# Cross-modality Attention Adapter: A Glioma Segmentation Fine-tuning Method for SAM Using Multimodal Brain MR Images

Xiaoyu Shi, Shurong Chai, Yinhao Li, Jingliang Cheng, Jie Bai, Guohua Zhao and Yen-Wei Chen*

**Abstract—** According to the 2021 World Health Organization (WHO) Classification scheme for gliomas, glioma segmentation is a very important basis for diagnosis and genotype prediction. In general, 3D multimodal brain MRI is an effective diagnostic tool. In the past decade, there has been an increase in the use of machine learning, particularly deep learning, for medical images processing. Thanks to the development of foundation models, models pre-trained with large-scale datasets have achieved better results on a variety of tasks. However, for medical images with small dataset sizes, deep learning methods struggle to achieve better results on real-world image datasets. In this paper, we propose a cross-modality attention adapter based on multimodal fusion to fine-tune the foundation model to accomplish the task of glioma segmentation in multimodal MRI brain images with better results. The effectiveness of the proposed method is validated via our private glioma data set from the First Affiliated Hospital of Zhengzhou University (FHZU) in Zhengzhou, China. Our proposed method is superior to current state-of-the-art methods with a Dice of 88.38% and Hausdorff distance of 10.64, thereby exhibiting a 4% increase in Dice to segment the glioma region for glioma treatment.

## I. INTRODUCTION

Brain tumors are classified as either primary or secondary brain tumors. Glioma is the most prevalent primary brain tumor [1]. Glioblastoma (GBM) is the most aggressive glioma. Note that <5% of patients survive five years after diagnosis [2]. According to the 2016 World Health Organization (WHO) Classification scheme for gliomas [3], glioma segmentation is a very important basis for genotype prediction and diagnosis. Therefore, it is necessary to predict glioma region for treating glioma. Magnetic resonance imaging (MRI) is a common method for helping doctors to do glioma diagnosis. Usually, MRI produces images in four modalities, which include T1, T2, T1CE, and Flair. Each modality has unique characteristics, which is useful in glioma region segmentation. Although it is difficult to segment glioma region using MRI images by an inexperienced person, it can be predicted with deep learning, which could help doctors save time in medical diagnosis. There have been many improvements achieved in the last few years in medical image segmentation using machine learning particularly deep learning. Olaf et al. [4] proposed a deep learning method using convolutional neural network (CNNs) with skip connection part to make a medical image segmentation. Furthermore, with the development of Vision Transformer (ViT) [5] in deep learning, Chen et al. [6] proposed a Tran-Unet and Hu et al. [7] proposed a Swin-Unet. With the development of devices, more and more foundation models such as GPT and Segment Anything (SAM) [8,9] are starting to emerge, which are trained using very large data sets and therefore have good results for different tasks. Because of the limited of difference between medical image data sets and real-world data sets, the deep learning-based method for medical data cannot just fine-tune the whole pre-trained foundation model. Therefore, there are already some methods for fine-tuning the foundation model for medical images [10,11]. These methods fine-tune the SAM by adding different adapters and achieve good results. However, due to the specificity of multimodal brain MRI, we cannot directly use these methods to process multimodal data. In this study, we proposed a cross-modality attention adapter for fine-tuning the foundation model (SAM) to improve the performance of glioma region segmentation task. We classify the modalities into T1-based modal and T2-based modal based on the modal characteristics, and then use two pre-trained foundation models to process these two modalities. We add the cross-modality attention adapter module to interact with the two modalities to fine-tune the model. During the training process, the backbone network is frozen and the parameters of the adapter as well as the mask decoder are trained. When comparing our model with the current state-of-the-art methods, our proposed method has better performance.

## II. METHOD

### A. Overview

The overview of the method is illustrated in Fig. 1. In our study, we propose a fine-tuning approach based on cross-modality attention adapter. First, based on the morphology of multimodal brain MRI images in medicine, we divided the four modalities into T1-based modality (consisting of T1 and T1ce) and T2-based modality (consisting of T2 and Flair). Since the original input of the SAM model is a real-world image with three channels of RGB, we use an initialized 1x1 convolutional layer to change the number of channels of the input from 2 to 3. After that, we use two pre-trained SAM image encoders to process the two modalities separately. In this process, we added the cross-modality attention adapter module to the ViT encoder block of the first n SAMs to interact the information between the two modalities. This process can be expressed using the following equation:

$$X_{T1}^{l+1} = C(F_{T1}^l, F_{T2}^l) + F_{T1}^l \qquad (1)$$

• Xiaoyu Shi, Shurong Chai, Yinhao Li and Yen-wei Chen are with the College of Information Science and Engineering, Ritsumeikan University, Shiga, Japan. E-mail: is0490sr@ritsumei.ac.jp

•Jingliang Cheng, Jie Bai and Guohua Zhao are with the affiliated hospital of Zhengzhou University, Zhengzhou, China.

•*Corresponding Authors: Yen-Wei Chen (chen@is.ritsumei.ac.jp)

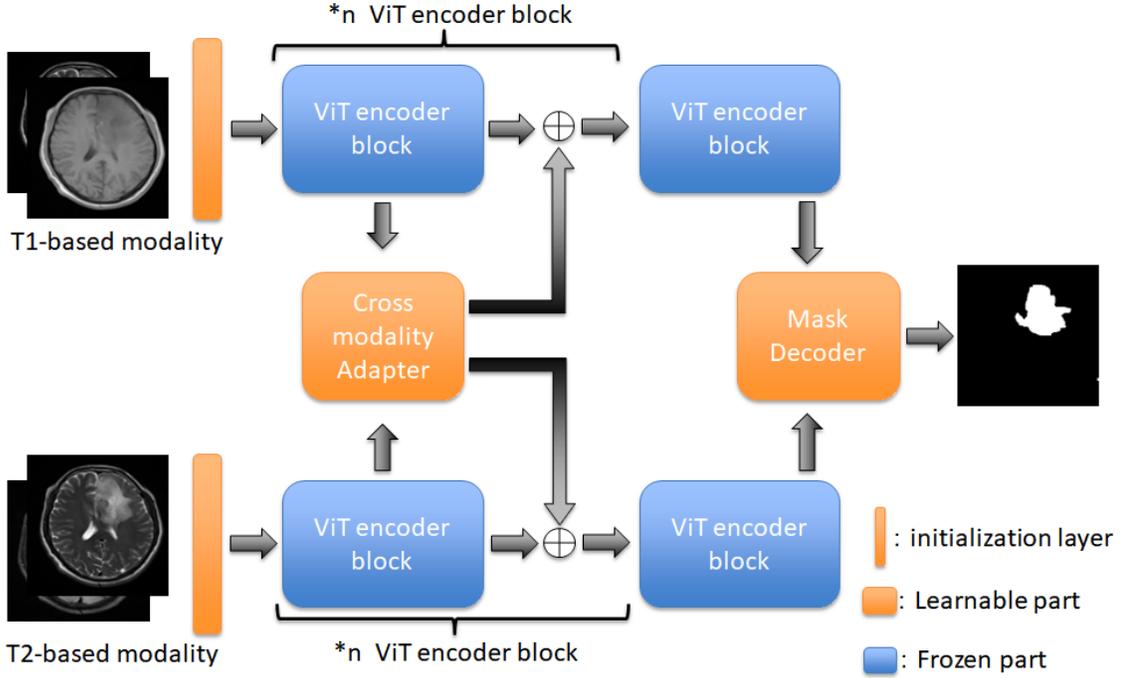

Figure 1. Overview of the proposed method

$$X_{T2}^{l+1} = C(F_{T1}^l, F_{T2}^l) + F_{T2}^l \quad (2),$$

where the $X^l$ is input of $l$ layer, C is our proposed cross-modality attention adapter module and the $F^l$ is the output features of $l$ layer in ViT encoder. Finally, we first sum the two high-level feature maps of two modalities and input them into the mask decoder and do not apply the prompt part to complete the segmentation of the glioma region. In the training process, we freeze all parameters of ViT encoder and train only cross-modality attention adapter and mask decoder. We use both dice loss and cross entropy loss to train the network, for which the loss function is calculated as follows:

$$L = \lambda_1 Dice(Pred, GT) + \lambda_2 CE(Pred, GT) \quad (3),$$

Where the Pred is the network prediction, the GT is the ground truth, and $\lambda_1, \lambda_2$ are the weights of two loss functions.

### B. Cross-modality attention adapter

The Foundation model works well for different downstream tasks, but fine-tuning the entire model is often ineffective due to equipment issues and differences between datasets. It is a common approach to use adapter to fine-tune the foundation model [10, 11]. In this study, we propose a special cross-modality attention adapter for multimodal brain MRI data. this method fine-tunes the whole model by interacting with different modality features to achieve better results. The cross-modality attention adapter is shown in Fig. 2.

After the same layer of ViT image encoder block, we interact features from T1-based modalities with features from T2-based modalities as a way to learn and fuse information from multimodal data. This process can be expressed by the following equation:

$$C = f(F_{T1}^l \oplus F_{T2}^l) \quad (2),$$

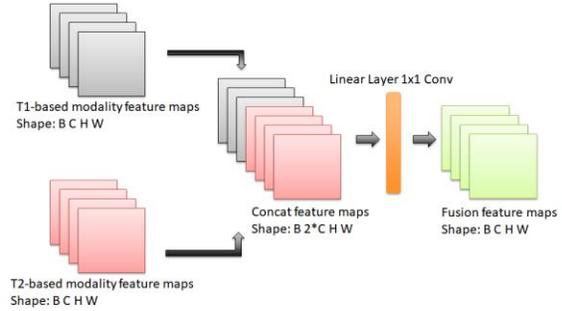

Figure 2. Cross-modality attention adapter

where the $F_{T1}^l, F_{T2}^l$ are the feature maps of two modalities, and the $f$ is a linear projection layer using 1x1 convolution to reduce the feature map channel from 2xc to c.

## III. EXPERIMENTS

### A. Data Set

In our experiments, 489 multimodality MRI images were obtained from the First Affiliated Hospital of Zhengzhou University in Zhengzhou, China. Each patient has four modalities of MR images (i.e., T1, T2, T1ce and flair images). The data for this study obtained from multiple imaging centers within the same hospital. The instruments, parameters, and conditions/circumstance utilized to acquire MRI images can vary substantially depending on the cases. This directly results in different spacing between cases and different modalities within individual cases. To solve this problem, we register all modalities to the T2 modality using the rigid registration method using Simple ITK tool [12,13]. Based on the imaging presentation, the physicians also labeled the abnormal areas of individual tumors as the ground truth of the

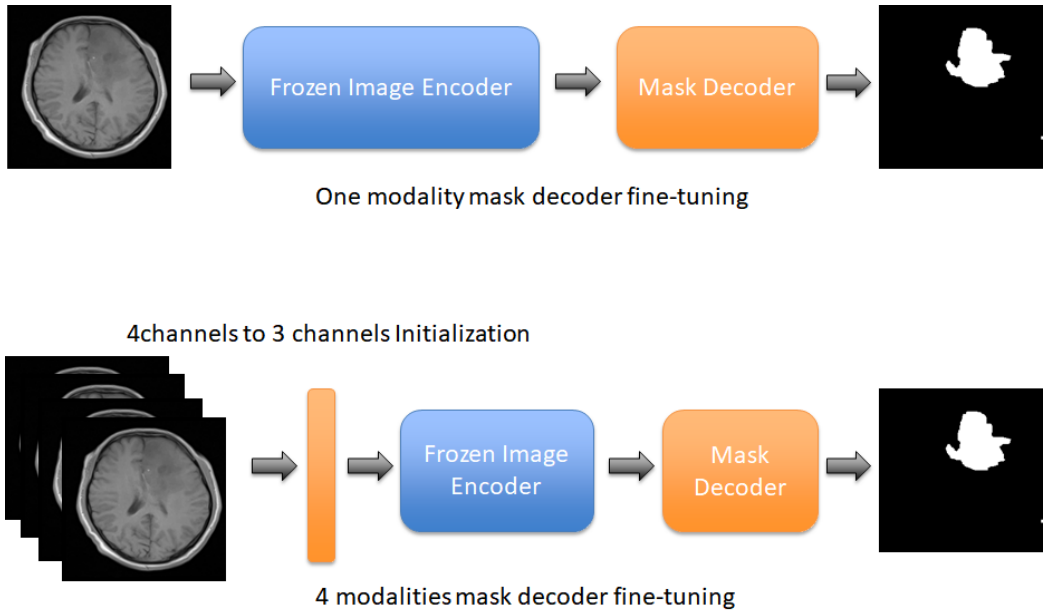
Figure 3. One modality test and early fusion test

experiments. We selected several slices from 3D MRI images to 2D images and divided the data sets to 729 images in training set and 245 images in test set randomly. The resolution of MR images used is 256×256.

*B. Experiments*

To confirm the influence of the cross-modality attention adapter for fine-tuning the SAM, we first test each modality and the 4 channels input as early fusion method performance. The networks and training methods are shown in Fig. 3. We selected Dice and Hausdorff distance 95 as our evaluation measures. The results of each modality and early fusion method are shown in Table 1. We also did the comparison experiments to the state-of-the-art methods. The results are shown in Table 2. Finally, we counted the total parameters of each model and the number of parameters used in training. The results of parameters are shown in Table 3.

Our method achieves better results compared to the individual modal and early fusion and SOTA methods. It is shown through the results that our proposed method fine-tunes the foundation model by interacting between the modalities and effectively improves the performance of the segmentation task. With adapter, we only trained a small fraction of the parameters (11%) to achieve better results, further demonstrating the effectiveness of our method.

We used the SAM ViTb model with first 6 layers' cross-modality attention adapter blocks, AdamW Optimizer with batch size 24, 0.001 learning rate, and data shuffle on one Nvidia GeForce RTX 3090 24GB GPU for training.

TABLE 1. COMPARISON OF EACH MODALITY

| Modality | Dice (%) | Hd95 |
|---|---|---|
| T1 | 68.96 | 24.13 |
| T1ce | 65.21 | 24.16 |
| T2 | 81.05 | 16.28 |
| Flair | 78.40 | 17.11 |
| Early fusion | 83.86 | 16.14 |
| **Our method** | **88.38** | **10.64** |

TABLE 2. COMPARISON WITH STATE-OF-THE-ART METHODS

| Method | Dice (%) | Hd95 |
|---|---|---|
| Trans-Unet [6] | 84.43 | 16.02 |
| Swin-Unet [7] | 84.05 | 16.25 |
| SAM [9] | 83.86 | 16.14 |
| **Our method** | **88.38** | **10.64** |

TABLE 3. COMPARISON OF THE PARAMETERS

| Method | All | Training | Percent |
|---|---|---|---|
| Trans-Unet [6] | 93M | 93M | 100% |
| Swin-Unet [7] | 43M | 43M | 100% |
| SAM [9] | 91M | 3.64M | 4% |
| **Our method** | **98M** | **10.80M** | **11%** |

IV. DISCUSSION AND CONCLUSION

In this study, we proposed a foundation model fine-tuning method using cross-modality attention adapter.

From the experimental results of our proposed method, we achieved Dice of 88.38, hd95 of 10.64, thereby obtaining better performance than the current state-of-the-art methods. In the applied research of computer medical diagnosis, multimodal information plays an important role. Because of the limitation of the data set of medical data, we must not only fine-tune the whole foundation model. In particular, in the task of segmentation of glioma, it is difficult to solve the problem of multimodal information fusion when fine-tuning the pre-trained foundation model. Therefore, we proposed a cross-modality attention adapter to interact features between two modalities, which finally achieved remarkable great improvement. In future, we intend to improve our methods, for our adapter and 3D images to achieve optimal performance in glioma region segmentation.

ACKNOWLEDGMENT

This work was supported in part by the Grant-in Aid for Scientific Research from the Japanese Ministry for Education, Science, Culture and Sports (MEXT) under the Grant No.20KK0234, No.20K21821, No. 21H03470 and No. 21K17774.